\documentclass[12pt]{article}
\usepackage{graphicx}

\usepackage{natbib}
\usepackage{wrapfig}

\newcommand{\be}{\begin{equation}}
\newcommand{\eeq}{\end{equation}}

\newcommand{\ba}{\begin{eqnarray}}
\newcommand{\ea}{\end{eqnarray}}
\newcommand{\om}{\omega}

\newcommand\eg{\textit{e.g.,\ }}

\newcommand{\Bf}{{magnetic field}}

\newcommand{\Ef}{{electric  field}}

\newcommand{\NS}{neutron star}
\newcommand{\NSs}{{neutron stars}}
\newcommand{\EM}{electromagnetic}
\newcommand{\BH}{{black hole}}
\newcommand{\BHs}{{black holes}}

\newcommand{\Fermi}{{\it Fermi}}

\newcommand\pubnumber{SNSN-323-63}
\newcommand\pubdate{\today}

\def\Title#1{\begin{center} {\Large #1 } \end{center}}
\def\Author#1{\begin{center}{ \sc #1} \end{center}}
\def\Address#1{\begin{center}{ \it #1} \end{center}}

\newcommand\pubblock{\rightline{\begin{tabular}{l} \pubnumber\\
         \pubdate  \end{tabular}}}
\newenvironment{Abstract}{\begin{quotation}  }{\end{quotation}}
\newenvironment{Presented}{\begin{quotation} \begin{center} 
             PRESENTED AT\end{center}\bigskip 
      \begin{center}\begin{large}}{\end{large}\end{center} \end{quotation}}

\def\beq{\begin{equation}}
\def\eeq#1{\label{#1}\end{equation}}
\def\eeqn{\end{equation}}

\def\beqa{\begin{eqnarray}}
\def\eeqa#1{\label{#1}\end{eqnarray}}
\def\eeqan{\end{eqnarray}}

\let\bar=\overbar

\def\eg{{\it e.g.}}

\def\Dslash{\not{\hbox{\kern-4pt $D$}}}
\def\dslash{\not{\hbox{\kern-2pt $\del$}}}

\def\msb{{\bar{\ssstyle M \kern -1pt S}}}

\begin{document}
\begin{titlepage}
\pubblock

\vfill
\Title{Afterglow jet breaks, GeV photons and the electromagnetic  model of short GRBs}
\vfill
\Author{ Maxim Lyutikov}
\Address{Department of Physics, Purdue University, 
 525 Northwestern Avenue,
West Lafayette, IN
47907-2036}
\vfill
\begin{Abstract}
We discuss three topics: (i) the dynamics of afterglow jet breaks; (ii) origin of  \Fermi-LAT photons; (iii) the \EM\ model of short GRBs
\end{Abstract}
\vfill

\begin{Presented}
GRB 2013, Huntsville Gamma-Ray Burst Symposium
\end{Presented}
\vfill
\end{titlepage}
\def\thefootnote{\fnsymbol{footnote}}
\setcounter{footnote}{0}

\section{Jet  breaks}

The concept of the light curve breaks is an important part of the theoretical studies of  GRBs
\cite{Rhoads99,Frail01}. We point out that the ``standard'' view point - the jet starts to expand laterally when the Lorentz factor becomes of the order of the opening angle  -  is flawed. At the afterglow stage, when the outflow launched by the central engine has transferred most of its energy to the forward shock, there  is no ``jet'', but a  non-spherical shock. Evolution  of strong non-spherical shocks is a  well studies  problems in fluid dynamics, in particular the evolution of non-spherical shock produced by nuclear explosions in the Earth atmosphere  \cite{Komp,LaumbachProbstein,ZeldovichRaizer,Bisnovatyi-KoganSilich95}.
For example, in  the  Kompaneets approximation the internal pressure of the
gas is assumed to be constant. Then the Rankin-Hugonio conditions
determine the normal velocity of the shock in the external inhomogeneous
medium.  A modification of 
the  Kompaneets approximation - a thin or snowplow shell approximation -
 has also been used extensively  \cite[\eg][]{1978ApJ...221...41W,1988ApJ...324..776M,1989Ap&SS.154..229B}. In a complimentary    Laumbach-Probstein approach  \citep {LaumbachProbstein} the streamlines of the shocked material
are assumed to be radial, thus   neglecting the 
lateral pressure forces.

These works,  especially that of  \cite{Komp}, bear an important lesson: post-shock pressure equilibration {\it does not mean spherical shock}. In the   Kompaneets approximation the post-shock pressure is constant, yet the shock is non-spherical. 

The relativistic generalization of the
  Kompaneets  and the Laumbach-Probstein methods have been discussed by  \cite{1979ApJ...233..831S,2012MNRAS.421..522L}. 
     Relativistic  dynamics provide extra support
for the thin shell method, since in the relativistic blast waves
the shocked material is concentrated in even narrower region
$R/\Gamma^2$ than in the non-relativistic Sedov solution.
In addition,  the limited causal connection (over the angle $\sim 1/\Gamma$)
provides a justification for the Laumbach-Probstein method on the
angle scale comparable to $1/\Gamma$.
As has been pointed out by  \cite{1979ApJ...233..831S}, the two methods - Kompaneets and
Laumbach-Probstein - become  very similar in the relativistic regime.

\cite{2012MNRAS.421..522L} has re-derive the relativistic Kompaneets equation  \cite{Komp,1979ApJ...233..831S} allowing for the arbitrary
velocity of the shock and arbitrary (angle-dependent) luminosity
and/or external density. It was found that the relativistic motion effectively
``freezes out'' the lateral dynamics of the shock front: 
only extremely strongly collimated shocks, with the opening 
angles $\Delta \theta \leq 1/\Gamma^2$, 
show  appreciable modification of profiles due to sideways expansion.
For less collimated profiles the propagation is nearly ballistic;    the sideways expansion of  relativistic shock 
becomes important only when they become mildly  relativistic.   

The origin of the  $\Delta \theta \sim 1/\Gamma^2$ scaling for fast lateral shock evolution is easy to understand: the  shock spreading relates to the {\it  rate} of change of the direction of the shock normal.  For  a typical collimated shock with  angle $\theta \sim  1/\Gamma$ this rate of change of the direction is  of the order of  the dynamic time scale {\it in the shock frame}: thus, it  is  $\Gamma$ times longer in the observer frame. In order to have a change in the direction of the order of unity on the dynamical time in the observer frame one need very fast, $t'_{dyn} \sim 1/\Gamma$, change in the shock frame - this then gives the scaling  theta $\sim1/\Gamma^2$ for quick sideways expansion in the observer frame.

These theoretical results are in general agreement with the detailed numerical simulations  \citep{2013ApJ...767..141V}, which demonstrate very slow, logarithmic with time, sideways expansion. The results discussed above  are all related to shock structure, while the observational implications are affected by other effects, like   limb brightening and subtle edge effects.

Thus, contrary to the commonly assumed fast  lateral expansion, the lateral evolution of the GRB shocks is effectively frozen out by the highly relativistic motion of the shock.  Only when the shock becomes weakly relativistic ti starts to evolve laterally on the dynamical time in the observer frame. This has important implications for the calculated  spectra of GRB afterglows. 

\section{GRBs and Fermi LAT photons: not synchrotron}

The detection of GRBs by the \Fermi\ satellite \citep{2009Sci...323.1688A} is an important probe of the GRB physics. The recent observation of the 95 GeV photon (125 GeV in rest frame) at $\sim 250$ seconds and 30 GeV  (40 GeV in rest frame)  at $\sim 35$ ksec  from GRB130427A can be used to {\it  exclude the synchrotron origin of LAT photons}. 
There is a {\it acceleration theory-independent}  upper limit on the frequency of synchrotron emission by radiation reaction-limited  acceleration of electrons \citep{2010MNRAS.405.1809L}. In astrophysics the effective  accelerating \Ef\ is a fraction $\eta \leq 1$ of the \Bf\ (this is equivalent to  acceleration on time scale of inverse cyclotron frequency $1/( \eta \om_{B,rel}) $, where $\om_{B,rel}=\gamma/\om_B$ is relativistic cyclotron frequency of a particle). Equating the acceleration rate and synchrotron energy losses,
\be
\eta e B c \sim  {e^2 \over c}  \gamma^2 \om_B^2
\end{equation}
the peak energy of  synchrotron emission is then
\be
\epsilon_{\rm max} ' \sim  \hbar  { m c^3 \over e^2} \approx 100 \mbox{ MeV}.
\label{emax}
\end{equation}
Note, the upper limit (\ref{emax}) assumes the most efficient, {\it non-stochastic}, DC-type acceleration. 

\begin{figure}[h!]
\includegraphics[width=\linewidth]{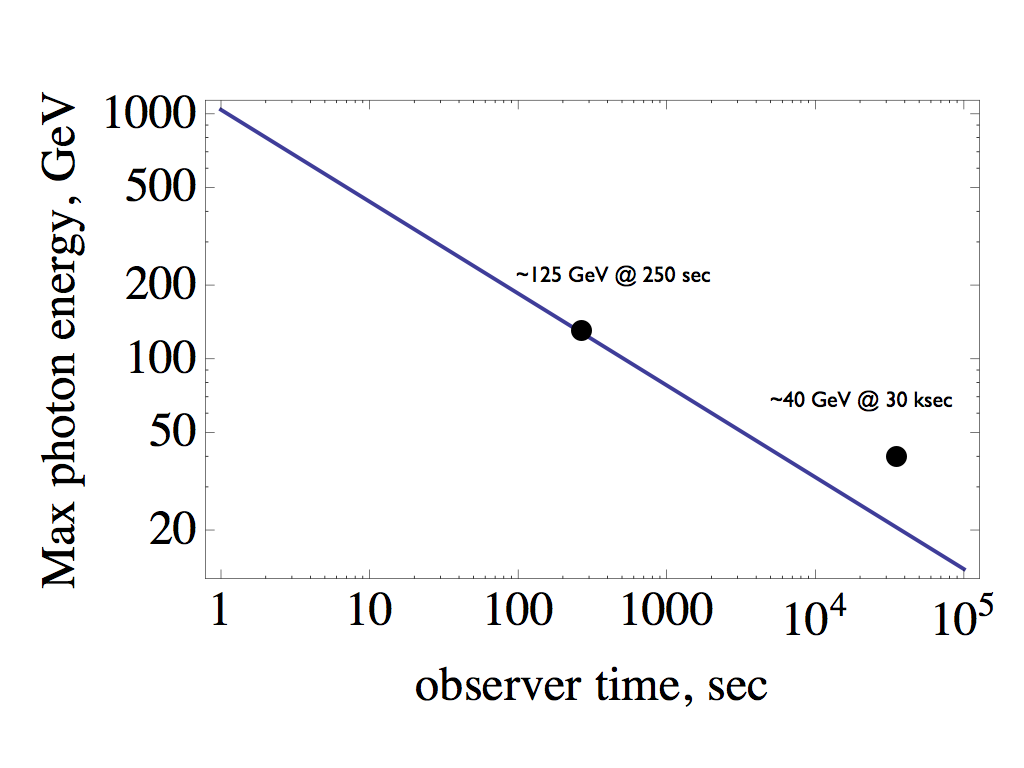}
\caption{Upper limit of synchrotron emission from the forward shock and two LAT photons from  GRB130427A. The very late high energy photon, $\sim 35 GeV$ at $\sim 35 $ ksec violates the upper limit on the synchrotron emission.}
\label{GRB-Upper-Sinch}
\end{figure}

If the emitting plasma moves with a Lorentz factor $\Gamma$ towards the observer, the observed maximal frequency is $\sim 2 \Gamma \epsilon_{\rm max}$. 
The Fermi photons come over times much longer than the duration of the prompt emission. Assuming the \cite{BlandfordMcKee} scaling of the Lorentz factor, we find
\be
\Gamma \sim\left(  {E_{iso} (1+z)\over c^5 t_{ob}^3 m_p n_{ISM} } \right)^{1/8}, \, \epsilon_{\rm max} = 2 \Gamma \epsilon_{\rm max} ' / (1+z)
\label{Emax}
\end{equation}
where we introduced the cosmological factor $(1+z)$.
The relation (\ref{Emax}) put  a constraint on the maximal synchrotron energy emitted at the observer time $t_{ob} $. The burst GRB130427A put particular tight constraints, see Fig. \ref{GRB-Upper-Sinch}. (We assumed $E_{iso} =10^{54}$ erg and $ n_{ISM} =1$.) Note that $ \epsilon_{\rm max} $ is very insensitive to the precise values of $E_{iso} $ and $ n_{ISM} =1$, $\epsilon_{\rm max}  \propto (E_{iso}/ n_{ISM} )^{1/8}$. 

\section{The electromagnetic model of short GRBs}
Many short GRBs show   long prompt tails that  last up to hundreds of seconds and may be energetically dominant over the initial spike \citep[by a factor of 30 as in GRB080503][]{Perley},. Such a long activity is problematic in the  model of  \NS-\NS\ mergers.   Numerical simulations indicate that the active stage of NS-NS coalescence typically takes  a very short time, 10 msec  \citep[\eg][]{Kiuchi}. 
  Highly spinning \NSs\ with hard equation of state may form a transient object which collapses  on 100 msec time scale \citep{2011PhRvD..83l4008H}, but {\it not} on hundred seconds time scale. 
 A small amount   of material, $\leq  10^{-3} M_\odot$,  may be ejected during the merger and accretes on time-scales of 1-10 secs, depending on the assumed $\alpha$ parameter of the disk \cite[\eg][]{Kiuchi, LiuNSNS,Faber}.

We developed an electromagnetic model of short GRBs  \citep{2013ApJ...768...63L} that explains  the two stages of the  energy release, the prompt spike and the prompt tail. The key point is the recent discovery that  an {\em  isolated}   black hole   formed from the collapse of a rotating neutron star can keep its open magnetic field  lines  for times much longer than the collapse time and, thus, can spin-down electromagnetically. 
The ``no hair'' theorem  \citep[][]{MTW}, a key result in General Relativity, states that an isolated  black hole  is defined  by  only  three parameters: mass, angular momentum, and  electric charge. We have recently demonstrate that  the ``no hair'' theorem  is not applicable for \BHs\ formed from collapse of a rotating neutron star \citep{2011PhRvD..83l4035L,2011PhRvD..84h4019L}.  
The key point in the classical  proof is that  the outside  medium is a vacuum. In contrast, the surroundings  of astrophysical high energy sources like pulsars and \BHs\ can rarely be treated as vacuum  \citep{GoldreichJulian,Blandford:1977,1992MNRAS.255...61M}.  

Merger of two neutron stars produces  a transient   supermassive fast  spinning  sheared neutron star in which  magnetic field may be amplified to $\sim  5 \times 10^{15}$ Gauss, typical of magnetars.  This magnetic field  extracts the rotational energy and drives an  electromagnetic wind that  may carry of the order of  $10^{50}$ ergs, limited by the collapse time of a  neutron star in the black hole. We identify the prompt spike in shorts GRBs as emission from highly relativistic Poynting flux-dominated wind produced by the transient supermassive neutron stars.

 The  black hole  resulting from the collapse of the supermassive  neutron star  releases its rotational energy  on a time scale of hundreds  to thousands  of  seconds. We identify the  prompt  tails  with the emission from the electromagnetic wind generated by the isolated highly spinning \BH. 
The corresponding powers and times scales  are sensitive functions of the neutron star masses, equation of state of the supermassive  neutron stars, and the level of magnetic field  amplification.  
The proposed model for short GRBs implies a different type of collimation of the outflow  than the conventionally envisioned jet-like structure: short GRBs produce equatorially, not axially, collimated relativistic outflow.

\begin{figure}[h!]
 \begin{center}
\includegraphics[width=.49\linewidth]{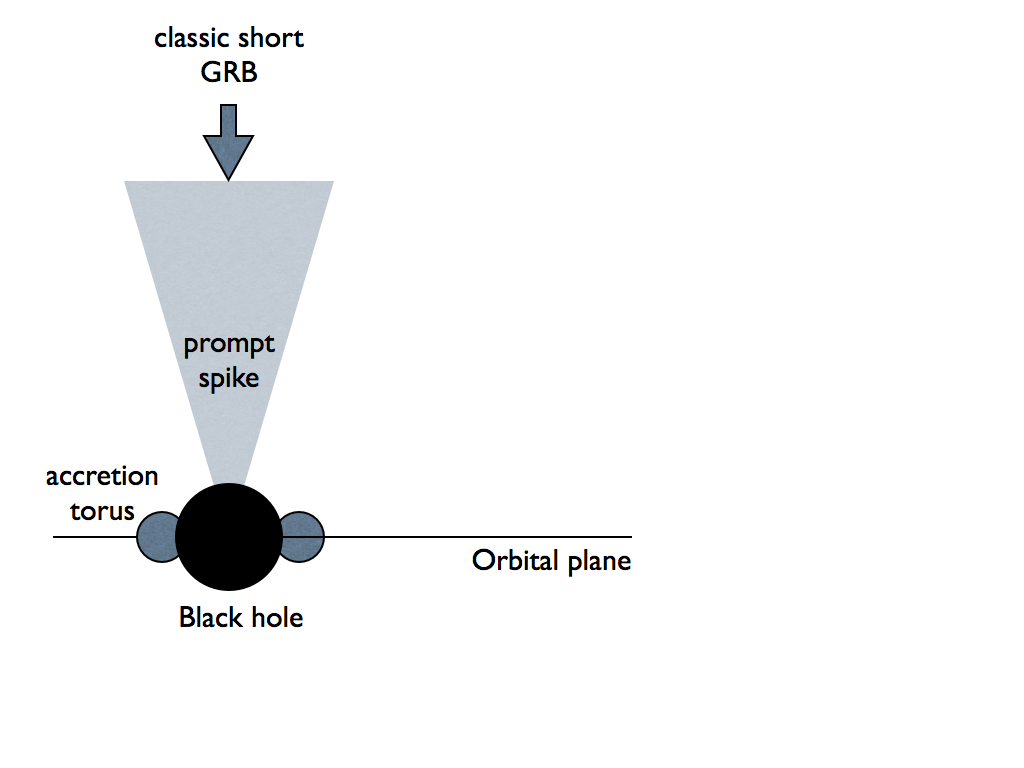}\includegraphics[width=.49\linewidth]{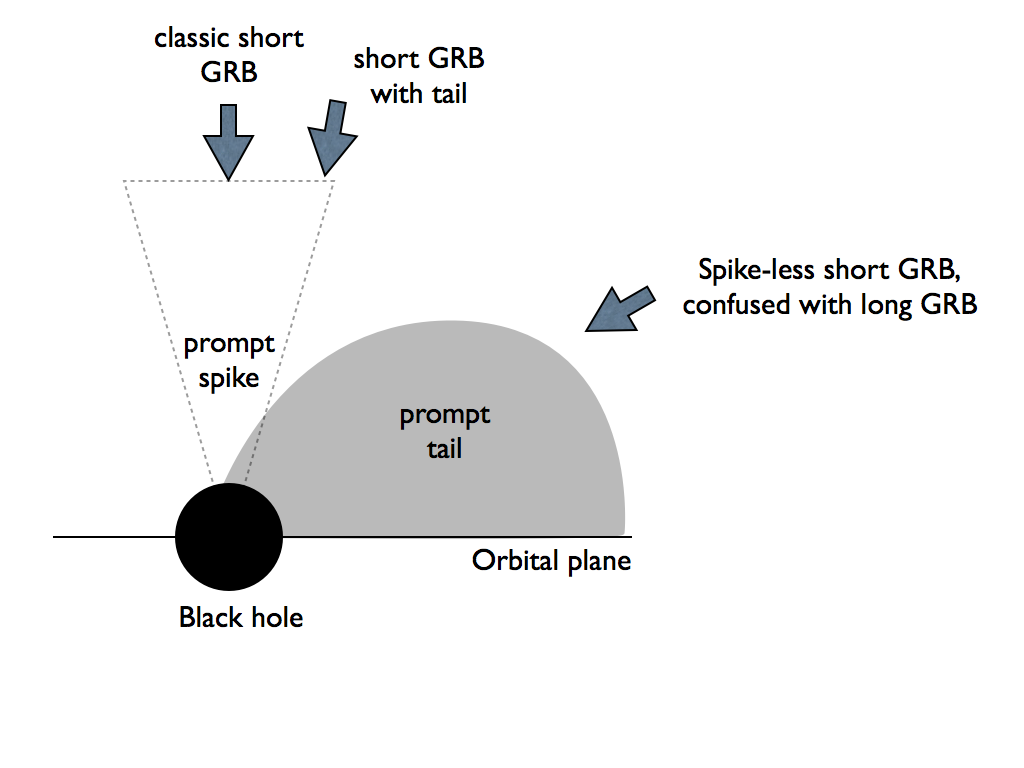}
\end{center}
\caption{The \EM\ model of short GRBs. At the prompt stage the surrounding torus collimates the BH-driven outflow along the rotational axis, producing a short GRB. After the torus is accreted, the isolated  BH spins down electromagnetically, producing equatorially-collimated outflow - the extended emission in short GRBs.  Equatorial observer may miss the prompt spike, mis-identifying the extended emission in short GRBs for a long, supernova-less GRB.}
\label{GRB-BH-Torus}
\end{figure}

\bibliographystyle{apj}
  \bibliography{/Users/maxim/Home/Research/BibTex}

\end{document}